\documentclass[a4paper,11pt]{article}
\usepackage[usenames,dvipsnames]{xcolor}
\usepackage{fullpage}
 \pdfoutput=1

\usepackage{amssymb,amsmath,amsthm,mathtools}
\usepackage{abbrevs}
\usepackage{longtable}
\usepackage{subcaption}
\usepackage{algorithm}
\usepackage{graphicx}

\usepackage{algpseudocode} 
\usepackage{pgfplots}

\usepackage{array}
\newcolumntype{P}{>{\raggedleft\arraybackslash}p{25pt}}
\newcolumntype{C}{p{10pt}}

\usepackage{tikz}
\usetikzlibrary{arrows}

\newname\fastcore{{\rm \textsc{fastcore}}}
\newname\fastcc{{\rm \textsc{fastcc}}}
\newname\findsparse{FindSparseMode}
\usepackage{cite}
\usepackage[labelfont=bf,labelsep=period,justification=raggedright]{caption}

\begin{document}

\begin{flushleft}
{\Large
\textbf{Fast reconstruction of compact context-specific metabolic networks via  integration of microarray data}}


Maria Pires Pacheco$^{1}$ and
Thomas Sauter$^{1}$
\\

\bf{1} Life Sciences Research Unit, University of Luxembourg, Luxembourg
$\ast$ E-mail: thomas.sauter@uni.lu
\end{flushleft}

%

\section{Introduction}
 Metabolism is a dynamic process that involves transport of metabolites and thousands of chemical reactions in which thousands of compounds are converted into others. Alternative pathways and branches are continuously activated or shut down to maximize metabolic efficiency in a specific context\cite{Segre2002}. Metabolism is so complex that the underlying processes can hardly be understood without using simplified mathematical representations. The most comprehensive formulations are genome-scale Reconstructions (GEMs).  For \textit{homo sapiens} there are several of these reconstructions like Recon 1 and 2 \cite{Duarte07, Thiele13} or the Edinburgh human metabolic network \cite{Ma2007/09/18/print}. Alongside these reconstructions follows the development of extensive reaction databases, like the HMR \cite{Agren12, agren2014identification} or HumanCyc \cite{romero2004computational, caspi2010metacyc}, which collect information to refine the available models. EMs and models derived from GEMs following omics data integrations were successively used to understand how perturbations in the metabolism lead to severe pathologies targets\cite{agren2014identification, mardinoglu2014genome, Folger11}.

  GEMs are generic representations of a cell of an organism comprising all the reactions that can potentially get active regardless of the environment and cell type. Therefore they do not cover the fact that the set of expressed genes and thereby the set of active reactions vary significantly in function of the cellular context.  This necessitates the generation of context-specific models containing only pathways predicted to be active in a given environment. Most context-specific reconstruction methods assume that the expression of genes correlates with the active state of the related reactions. Although, this assumption is only partially justified, context-specific models showed a higher predictive power than the GEMs from which they were derived from~\cite{Jerby10,Becker08}. This is due to GEMs containing multiple alternative pathways that are rarely simultaneously active and therefore tend to have an increased number of false negatives in gene essentiality assays compared to context-specific models that only comprise the active alternative pathway~\cite{Folger11,hyduke2013analysis,bordbar10}.

 Recently we proposed an algorithm for the fast reconstruction of compact context-specific metabolic networks (FASTCORE) that allowed dropping the reconstruction time of context-specific networks to the time order of seconds \cite{Vlassis2014}. This extremely low computational demand opens new possibilities for improving the quality of the models. Several rounds of model reconstruction, testing of the model’s predictions against real experimental data, curation steps of the input model and the set of core reactions as well as cross-validations assays are required to reconstruct high-quality models. These semi-automated model curations steps are in such extend not possible with competing algorithms due to their high computational demands. 
  FASTCORE requires as input a GEM and a set of core reactions being active in the context of interest. FASTCORE identifies a close to minimal set of non-core reactions from the input model to be added to the core set in order to obtain a consistent model, in which every reaction in the model is able to carry a non-zero flux. To reconstruct compact models, the inclusion of non-core reactions is penalized \cite{ Vlassis2014}. 
  
 But the question which genes are expressed in a given cell-type and therefore the establishment of the core set of reactions is non-trivial. Microarray expression data, so far the most popular data source for model contextualization, allow comparing the probe expression levels between two conditions. But probe effects do not allow a direct comparison between the probe sets, as the amount of noise is non-negligible and varies within probe sets so that higher intensity does not imply higher expression levels or even to state that a gene is expressed\cite{ Zilliox07}. Further the intensities retrieved from microarray data are continuous while FASTCORE, like most context-specific algorithm require a binary input data (the establishing core set).
 
  To adapt FASTCORE for the integration of microarray data, we therefore propose a new workflow: FASTCORMICS. FASTCORMICS requires as input microarray data and a GEM. Like FASTCORE, FASTCORMICS is devoid of heuristic parameter settings and has a low computational demand with overall building times in the order of a few minutes  FASTCORMICS preprocesses the microarrays data with the discretization tool Barcode\cite{Zilliox07, mccall2011gene}. Barcode uses prior knowledge on the intensity distribution of each probe set for a given microarray platform to segregate between expressed genes and non-expressed genes. The preprocessing step with Barcode allows circumventing the need of setting a heuristic expression threshold that segregates between expressed and non-expressed genes as e.g. in \cite{Becker08,Zur10, Folger11}. Choosing such a threshold is arbitrary and critical for the output metabolic models as in response to this threshold complete branches, alternative pathways, or subsystems might be included or excluded, thereby heavily changing the functionalities of the model. Further, Barcode shows a better correlation between predicted expression and protein expression than competing discretization methods for the segregation of gene expression and allows to reduce batch and lab-effects that affect measurements \cite{Zilliox07}.

 FASTCORMICS was validated via an essentiality assay performed on two cancer models (cancer1 and cancer2) extracted respectively from Recon 1 and Recon 2 by the FASTCORMICS workflow. The predicted essential genes were compared to a list of essential genes established in \cite{luo2008highly} in a shRNA knockdown screen on cancer cell lines. The predictive power of the reconstructed context-specific models was compared to the original GEMs and to a generic cancer model \cite{Folger11} built by the competing MBA algorithm\cite{Jerby10}. Furthermore, as a second quality control step, enrichment in neoplasia-related genes retrieved from the DisGeNET database in the predicted essential genes was assessed with a hypergeometric test to check for the model ability to predict cancer relevant genes. In general, FASTCORMICS outperforms competing algorithms and allows obtaining high-quality, robust models in a high-throughput manner. This will allow the use of metabolic modelling as routine process for the analysis of microarray data e.g. in the field of personalized medicine.

 \section{Materials and methods}
 
 \subsection{ Workflow overview}
 \begin{figure}[!htbp]
 \centering

       \includegraphics[width=0.5\textwidth]{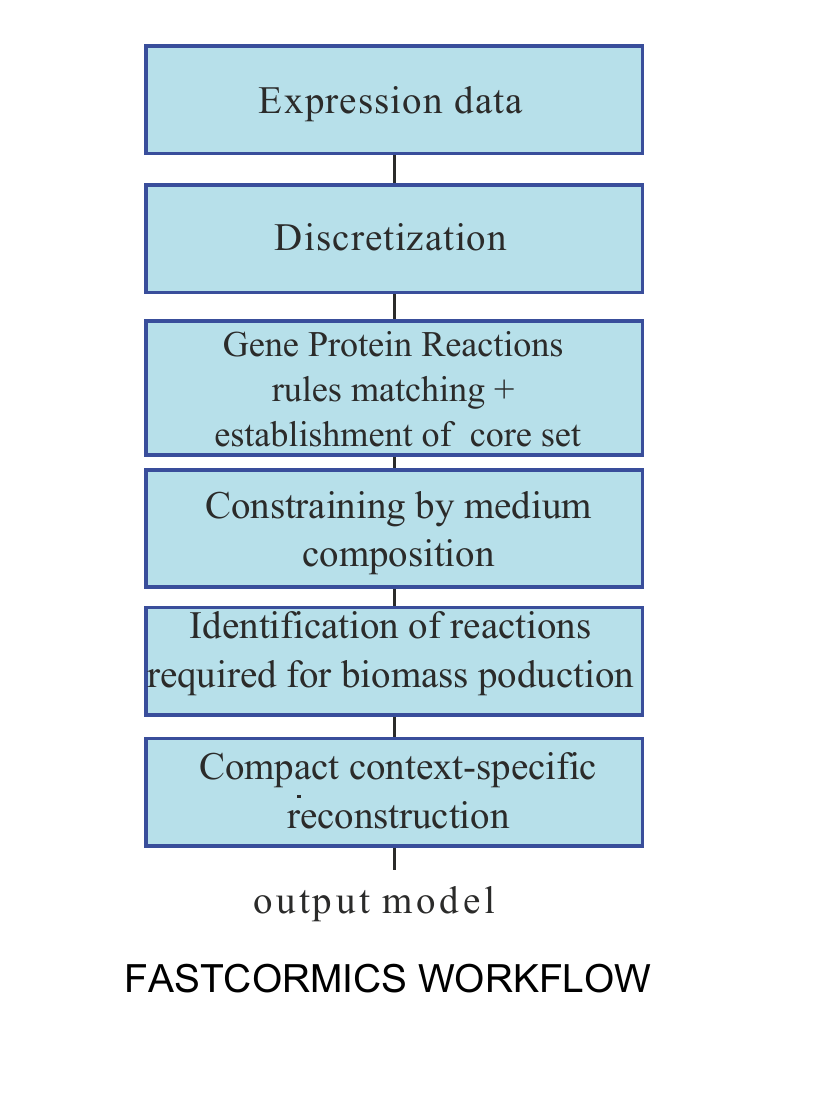}
         \caption{\textbf{FASTCORMICS Workflow:} After discretization of the microarray data with Barcode, the expressed genes are mapped to the input model according to the Gene-Protein-Reactions rules . The FASTCORE core set is composed of reactions under the control of Barcode-supported genes. Optionally, the model can be constrained in function of the medium composition and a biomass function. A modified version of FASTCORE, that allows the definition of a set of non-penalized reactions (in this study: barcode-supported core reactions) is run. The modified version of FASTCORE forces the biomass function to carry a non-zero flux while penalizing the inclusion of non-core reactions. The output of the modified FASTCORE is then added to the core set and the modified FASTCORE is run again, this time, forcing all core reactions to carry a flux while penalizing non-core reactions. Transporters are removed from the core set, but are not penalized as explained in the main text. }
          \label{fig:workflow}
 \end{figure}
 The general workflow of FASTCORMICS (figure1) contains a discretization step with Barcode to obtain a list of genes expressed in the context of interest. The latter is then mapped to the consistent generic model via the model's Gene-Protein-Reactions Rules (GPR) to obtain a list of active reactions (core reactions).  For reactions that are under the control of one gene, the discretized gene expression value is directly mapped to the reaction; otherwise if more genes are associated to a reaction, the relationship between the genes and the reaction is given by Boolean Rules. A Boolean AND means that all the genes have to be expressed to activate the reaction, which is typically the case when a reaction is controlled by a complex of proteins. Therefore the minimum of the discretized values is mapped to the reaction. A Boolean OR signifies that only one gene has to be expressed, the maximal value is mapped to the reaction.  Boolean ANDs and ORs can be combined inside a same rule i.e.  ((A AND B) OR C), in this example the minimal value D is computed between A and B then the maximum between D and C is matched to the reaction.  Reactions that are predicted active according to the GPR rules constitute the set of core reactions that are fed into a modified version of FASTCORE (mFC) that allows leaving a set of reactions not penalized besides defining core and non-core reactions. The inclusion of the set of non-penalized reactions are, unlike core reactions, not forced but only preferred over the inclusion of non-core reactions, which are penalized. Barcode-supported transporters, are a good example for the need of this new reaction set. Transporter reactions are  generally under the control of  promiscuous genes (in the consistent version of Recon 2\cite{Thiele13}, e.g. the gene SLC7A6  controls 294 reactions) and therefore transporters should be removed from the core set as otherwise whole subsystems would be included in the output model due to one gene. Nevertheless, the inclusion of barcode-supported genes should be preferred over non-core reaction which are not supported and therefore barcode-supported transporters are not be penalized. For more details on FASTCORE see the original paper \cite{Vlassis2014}. A Matlab implementation of the FASTCORE algorithm can be downloaded from bio.uni.lu/systems\_biology/software. 
 Two optional steps can be included in the workflow. The first one allows to constrain the model with respect to the medium composition if this is information is available. The uptake reactions for metabolites that are not present in the medium are shut down and FASTCC \cite{Vlassis2014} is run to remove reactions that cannot carry a flux due to these medium constraints. The second optional step allows adding a biomass function to the model. FASTCORMICS forces the biomass function to carry a flux while penalizing the inclusion of non-core reactions (Figure 1). Core reactions, including core transporters are not penalized in order to find, within the different alternatives sets of reactions that allows the production of biomass, the one that contain the highest number of core reactions. The output reactions of the modified FASTCORE are then added to the core set and the modified FASTCORE is run a second time  to now force all the core reactions to carry a flux while penalizing the non-core reactions. Transporter reactions are removed from the core set but are not penalized during the reconstruction to favour barcode-supported transporters over non-core reactions that are not supported. If no biomass function was added, Fastcormics is only run ones.

 \subsection{Microarray preprocessing with Barcode of cancer cell line data}

 The NCI dataset composed of 174 Hgu133plus2 arrays corresponding to 59 cancer cell lines was downloaded from the Cell miner web page \cite{shankavaram2009cellminer} and read in  version 2.15.1 of  R   with the affy package (1.36.1). The arrays were normalized with the frozen Robust Multi-array Average package (fRMA version 1.14.0)\cite{mccall2010frozen} and then processed with the Barcode \cite{Zilliox07} function using the hgu133plus2frmavrecs vector (version 1.1.12) into a list of expressed genes and another list of genes with intensity values not significantly different from the intensities obtained for same probesets in an unexpressed state (Figure 1). The ubiquity of expression (number of arrays for which a gene is expressed over the total number of arrays) was computed for each gene and a list of genes Entrez IDs with their respective score was then loaded in Matlab (version 2013a) and mapped via the Gene Protein Reactions Rules (GPR) to the consistent version of Recon1 (consistRecon1, 2469) and Recon2 (consistRecon2, 5317) obtained with FASTCC . Reactions tagged as expressed in ≥90\% of the 174 arrays were included in the core set with the exception of Barcode-supported transport reactions (core transporters), as the genes controlling the latter are in general promiscuous.  The core transporters were excluded from the core set, but were not penalized later during the building process. 
 
 \subsection{Building of cancer models constraint to growth on RPMI medium}
  To simulate the growth of the cancer cells on RPMI medium, the uptake reactions of the consistent versions of Recon 1 and Recon 2 were first constrained with respect to the medium composition and a biomass function taken from \cite{Wang12} was added to the GEMs. FASTCC \cite{Vlassis2014} was run to remove reactions that are not able to carry a flux due to these additional medium constraints (Figure 1).
  
   The modified FASTCORE was then run on the medium-constrained models forcing the biomass function to carry a flux while penalizing the inclusion of non-core reactions. This step allows selecting preferentially among all the alternative pathways for the production of biomass the one with the lowest number of non-core reactions, not penalizing the inclusion of core reactions. The reactions required to allow a biomass production were then added to the core set and the modified FASTCORE was run again now forcing the inclusion of all core reaction while penalizing the non-core reactions with the exception of core transporters (Figure 1). 
   
 \subsection{Model validation based on a knock-out experiment to identify essential genes}

  A knock-out experiment was performed on the obtained cancer models as described by \cite{Folger11} applying Flux Balance Analysis (FBA)\cite{Orth11}. In \cite{Folger11}, a gene is considered essential if its knock-downs results in a decrease of the growth rate of more than 1\%. To allow, a comparision with \cite{Folger11}, the 1\% criteria was taken over, but the experiments were also repeated with a growth rate dicrease criteria of 50\% (growth -50\%). The lists of essential genes were compared to the ranked list of 8000 genes established by \cite{luo2008highly} based on a shRNA knockdown screen on cancer cell lines. The rank of essential genes were compared to the rank remaining metabolic genes (set of genes associated to Recon2 minus the essential genes) with a Kolmogorov-Smirnov test (KS-test). In addition 1000000 random sets of genes of the same size were created and the respective KS-test was computed for evaluating the likelihood to obtain the same or better KS-score by chance. 
 To further validate the predicted essential genes, a list of neoplasia-related genes was retrieved from DisGeNET\cite{queralt2013disgenet}, a database for gene-disease associations. A hypergeometric test was performed to evaluate the enrichment of neoplasia-related genes in the predicted essential genes.

 \section*{Results}
 Two generic cancer models were obtained via the integration of microarray data from the NCI dataset\cite{shankavaram2009cellminer} with the FASTCORMICS workflow. The first model (cancer1), derived from Recon 1, is composed of 816 reactions (Table 1) and is therefore bigger than the cancer model derived in \cite{Folger11} (772 reactions).  The second model (cancer2) was extracted from Recon 2 and is composed of 1332 reactions. Essentiality assays performed on cancer1 and cancer2 predict 188 and 106 essential genes, respectively. The lists of essential genes were compared to a ranked list of 8000 genes established by \cite{luo2008highly} via a shRNA knock-outs assay. Metabolic genes are slightly overrepresented in the top of the list (data not shown), suggesting that metabolic genes of Recon 1 and Recon 2 are more essential than the remaining genes on the list. Furthermore, the distribution of essential genes of the cancer models is different from the remaining metabolic genes and shifted towards the top of the ranked list as showed by a one-side KS-test giving p-values of 7.7232e-04 and 0.0095 for cancer1 and cancer2, respectively, compared to a p-value of 0.0284 for the cancer model published in \cite{Folger11} that was built by the MBA algorithm\cite{Jerby10}.
 \begin{table}
 \caption{Comparison of the essential genes found by the in silico essentiality assay to a ranked gene list established by \cite{luo2008highly} defined as the effect of shRNA knock-downs on the proliferation of cancer cells. In \cite{Folger11}, a gene is considered essential if its knock-downs resulted in a decrease of the growth rate of more than 1\%. To allow, a comparision with \cite{Folger11}, the 1\% criteria was taken over, but the experiments were also repeated with a growth rate decrease criteria of 50\% (growth -50\%).  *The number of essential genes was taken from supplementary file msb201135-sup-0002.xls of \cite{Folger11} }
 \label{fig:KO}
 \begin{tabular} { p{4cm}||l|l| l| l| l|}

 output model & generic model & size & essential genes &   p-value &   permutation p-value\\ 
  \hline cancer folger
   & Recon1  & 772 & 178* &   0.0284 & 0.0059     \\ 
  \hline medium constrained Recon 1 + biomass  & Recon 1&  1922 &  78 &  0.1908  & 0.1444\\
  \hline medium constrained Recon 2 + biomass  & Recon 2 & 4246 & 36 &  0.7777 & 0.7398    \\ 
 
  \hline
  \hline  cancer1  & Recon1 & 816 &  188  &  7.7232e-04 &1e-06  \\ 
  \hline cancer1 (growth -50 \%) & Recon1 & 816 & 92 & 0.0489 &  0.0324 \\
  \hline  cancer2 &  Recon2  & 1332&  106 &  0.0095 & 0.0022   \\ 
 
 \end{tabular}
 \end{table}

 \begin{table}
 \caption{Hypergeometric test quantifying the enrichment of neoplasia-related genes retrieved from DisGeNet\cite{queralt2013disgenet} in the set of essential genes , a database of disease-gene associations. In \cite{Folger11}, a gene is considered essential if its knock-downs resulted in a decrease of the growth rate of more than 1\%. To allow, a comparision with \cite{Folger11}, the 1\% criteria was taken over, but the experiments were also repeated with a growth rate dicrease criteria of 50\% (growth -50\%).  } 
 \begin{tabular}{ p{4cm}|| p{2.5cm}| p{1.8cm}|p{3cm}| p {2.5 cm}| c|c| c|}
 output model & essential genes (EG) &  EG in DisGeNet &   genes in the generic models (GG) & GG in DisGeNet  & p-value \\ 
  \hline cancer folger &  178 &  84 & 1168 & 449 & 4.6 e-06     \\ 
   \hline medium contrained Recon1 + biomass  &  78 & 33  & 1168 & 377 &  0.0350\\
  \hline
    medium  constrained Recon2 + biomass &  36 &  14 & 1599   & 433 & 0.0806   \\

 \hline  cancer1 & 188 & 90 & 1168 & 377   &  8.1e-07  \\ 
 \hline  cancer1 (growth -50\%) & 92 & 39 & 1168 & 377   &  0.0063  \\ 
 \hline  cancer2 &  106 & 45  &  1599& 433&   2.9e-04    \\ 
  \hline
  
  \hline

 \end{tabular}
 \end{table}

  A permutation test (Table 1) showed that the likelihood of finding a gene set of the same size with a  better KS-score by chance is low with a p-value of 1e-6 and 0.0022 for cancer1 and cancer2, respectively, against a p-value of 0.0044 for the cancer model built by the MBA algorithm \cite{Folger11}.
 
   The Recon1 and Recon 2 models, further constrained by the medium composition, allowed to only identify a smaller set of essential genes (Table 1) and their distribution were not significantly different from the distribution of the metabolic genes for Recon1 and Recon 2 , confirming that context-specific models perform better than the generic  genome-wide reconstructions from which they are extracted.  
 
 The hypergeometric test (Table 2) showed that the neoplasia-associated genes retrieved from the DisGeNet database\cite{queralt2013disgenet} are over-represented in the essential genes for all models This confirms that the essential genes predicted by the cancer models are not false positives due to model-specific bias like the lack of alternative pathways in the generic model or the removal of the latter due to a  high threshold

 \section*{Discussion}
 
 We extended the FASTCORE algorithm \cite{Vlassis2014}, that allows reconstructing compact context-specific metabolic models in the time order of seconds, towards the integration of microarray data. The resulting FASTCORMICS workflow, that performs a reconstruction in the order of a few minutes, was validated by an essentiality test performed on generic cancer models, built by FASTCORMICS following the integration of an NCI microarray dataset. The essential genes predicted by FASTCORMICS ranked highly in the list of essential genes establish by \cite{luo2008highly} based on the shRNA knockdown effects on cancer cells proliferation and also were enriched for neoplasia-related genes. 
 
   Unlike most competing algorithms, FASTCORMICS does not depend on the introduction of heuristic thresholds for the segregation of expressed and non-expressed genes, which turns the models built by FASTCORMICS more robust and less prompted to over-fitting of the data. For the cancer model built by the competing MBA algorithm, considered as state of the art for context-specific metabolic reconstructions, the threshold was set at an intensity value of 200\cite{Folger11}. The size of the output model and therefore the number of essential genes is very sensitive to the choice of this threshold. A higher threshold would have led to overestimation of essential genes, due to a reduced number of alternative pathways included in the model, whereas a lower threshold would have led to an underestimation of the number of essential genes. Moreover, as the intensity distribution varies between experiments and platforms, the value of 200 that seems adequate for this dataset, might be inadequate for another one. Further, when comparing cancer1 with the cancer model built by MBA algorithm\cite{Jerby10}, cancer1 performed slightly better on the essentiality test and the enrichment test.
   Finally, like for FASTCORE, core reactions fed to MBA must have a high confidence level as all core reactions will be included in the output model. Core reactions with low confidence level, due to a too low threshold or lab effects, can cause the inclusion of a great number of non-core reactions into the output model in order to guarantee a flux trough all core reactions. Barcode\cite{Zilliox07} used with the default setting (as performed here) requires five standard deviation above its null mean to be regarded as expressed which  drastically limits the source of error in establishing the core reactions set.
 Other competing algorithms using omics data for building of context-specific model like GIMME \cite{Becker08}, IMAT\cite{Zur10} or, mCADRE\cite{Wang12}, INIT\cite{Agren12} or the MBA algorithm \cite{Jerby10} have higher computational demands due to the used mixed integer linear programming, and/or require setting of one expression, respectively two expression thresholds for IMAT\cite{Zur10}. FASTCORMICS outperforms its competitors due to its robustness and the low computational demand due to the efficient use of linear programming.

 \section*{Acknowledgements}
  We would like to thank Thomas Pfau and Lasse Sinkkonen for their feedback and for the interesting discussions.
 
 The research was funded by the Life Sciences Research Unit, University of Luxembourg. MPP was supported by a fellowship from the National Research Foundation of Luxembourg (FNR; http://www.fnr.lu) (AFR 6041230). The funders had no role in study design, data collection and analysis, decision to publish, or preparation of the manuscript.

 \bibliographystyle{plain}
 \bibliography{Maria2}
 
 \end{document}